\documentclass{article}

\usepackage{amssymb}
\usepackage{graphicx}
\usepackage{subcaption}
\usepackage{url}
\usepackage{bm}
\usepackage{xcolor}
\usepackage[version=3]{mhchem}
\usepackage{placeins}
\newcommand*\rev[1]{\textcolor{black}{#1}}

\usepackage{times}
\usepackage{chemformula}
\usepackage{textcomp}
\usepackage{xspace}

\usepackage{booktabs} 
\usepackage{dcolumn}
\newcolumntype{d}[1]{D{.}{.}{#1}}

\usepackage{multirow}
\usepackage{afterpage}
\usepackage{blindtext}
\usepackage{longtable}
\usepackage{hhline}

\usepackage{xcolor}
\usepackage{soul}
\definecolor{darkorange}{rgb}{1.0, 0.55, 0.0}

\date{}

\begin{document} 
\begin{center}
{\LARGE Evidence for Phenylium Reactivity under Interstellar Relevant Conditions
}\\ 
\end{center}
Jean-Christophe Loison$^{1}$,
Corentin Rossi$^2$,
Nicolas Solem$^{3,4}$
Roland Thissen$^{3,4}$,
Claire Romanzin$^{3,4}$,
Christian Alcaraz$^{3,4}$,
Ugo	Jacovella$^{2,\ast}$

\small{

\noindent 
$^{1}$Institut des Sciences Moléculaires, CNRS, Université de Bordeaux, F-33400 Talence, France\\ 
$^{2}$Universit\'e Paris-Saclay, CNRS, Institut des Sciences Mol\'eculaires d'Orsay, 91405 Orsay, France\\
$^3$Universit\'e Paris-Saclay, CNRS, Institut de Chimie Physique, UMR8000, 91405 Orsay, France \\
$^4$Synchrotron SOLEIL, L’Orme des Merisiers, 91192 Saint Aubin, Gif-sur-Yvette, France\\
$^\ast$ Corresponding author. E-mail: ugo.jacovella@cnrs.fr
}

\baselineskip24pt





\rev{Recent work by Kocheril \textit{et al.}\cite{kocheril2025} has questioned the role of phenylium (\ce{C6H5+}) in interstellar aromatic chemistry. It is reported that cyclic \ce{C6H5+} is unreactive toward key interstellar molecules, including \ce{H2} and \ce{C2H2}, challenging its proposed role in polycyclic aromatic hydrocarbon (PAH) formation. Here, we present experimental and theoretical evidence demonstrating that phenylium reacts efficiently with \ce{C2H2} under astrophysically relevant conditions through a barrierless mechanism. These findings reinstate phenylium as a crucial precursor in the formation of the first aromatic ring, a critical bottleneck in models of PAH growth in space.}


Computationally, we examined the encounter of \ce{C2H2} with \ce{C6H5+} at the M06-2X/AVTZ level. Figure~1a reveals that the entrance channel is barrierless. The ground-state singlet potential energy surface, including zero-point energy corrections is presented in Figure~1b (See Supplementary Table 1 for details). Although phenylium retains aromatic character, the formation of a bicyclic adduct (\textbf{1}) does not disrupt the $\pi$ system \cite{Nicolaides1997}. The reaction involves two sigma bonds between the non-hydrogenated carbon of \ce{C6H5+} and the acetylene carbons, similar to the bonding in \ce{C6H7+}, which also retains aromaticity. The adduct may isomerize into \ce{C6H5CCH2+} (\textbf{2}) or a five-membered bicyclic species (\textbf{3}). These can subsequently evolve through H-loss into three exothermic products (\textbf{i}, \textbf{ii}, and \textbf{iii}). Among these, only the formation of a fused pentalene-type ring (\textbf{iii}) is slightly exothermic.

\begin{figure}[!t]
	\includegraphics[width=1\linewidth]{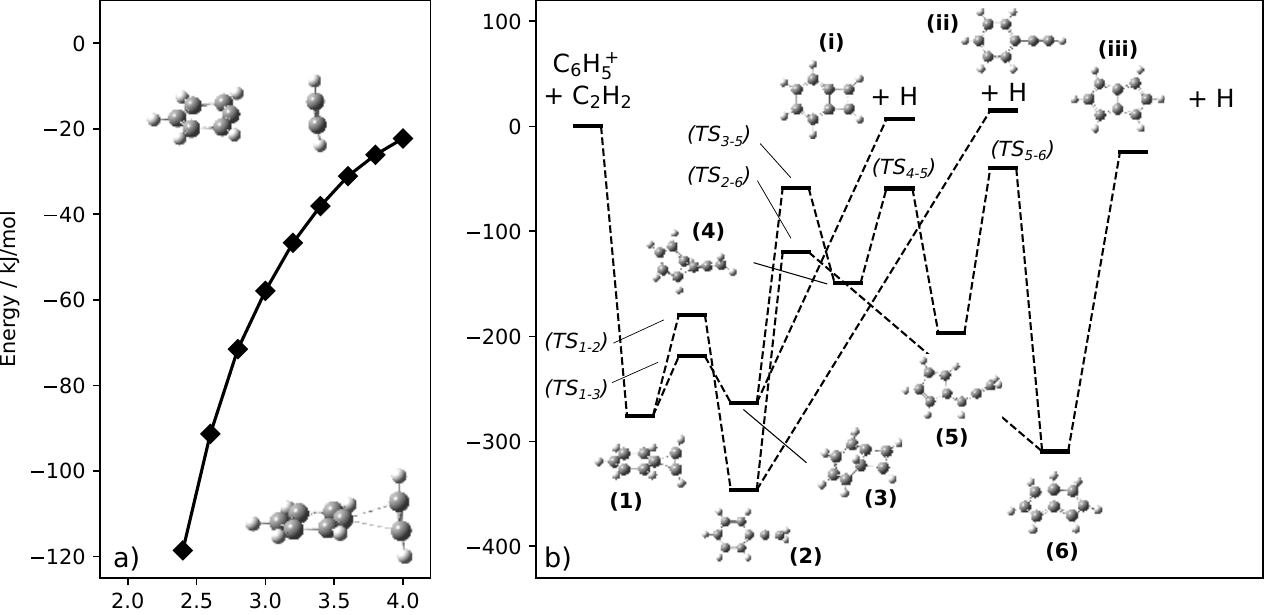}
	\caption{a) Energy pathway for \ce{C2H2} approaching phenylium, with all coordinates optimized except the distances between the acetylene carbons and the non-hydrogen-bearing carbon of phenylium. Phenylium dihedral angles were fixed at their isolated values. b) Potential energy surface for the \ce{C6H5+} + \ce{C2H2} reaction (energies given in Supplementary Table 1).} 
	\label{calcC2H2}
\end{figure}

We also carried out experimental work investigating the reactivity of phenylium with acetylene. Phenylium was generated via one-photon vacuum ultraviolet (VUV) dissociative ionization of nitrobenzene, a process known to yield exclusively the phenylium structure near threshold, as previously demonstrated by Cooper \textit{et al.} and references therein.\cite{cooper2001} According to the state-of-the-art Active Thermochemical Tables (ATcT) database \cite{ATcT}, the appearance energy (AE) of phenylium is 11.49 $\pm$ 0.02 eV at 300 K and 11.44 $\pm$ 0.02 eV at 0 K. The accurate determination of AEs has historically been experimentally challenging, therefore, we adopted the ATcT values. Although we recorded data at photon energies as low as 11 eV, the first reaction products were observed only at 11.40 eV, consistent with the predicted AE. Figure~2b) shows cross sections and rate constants for the \ce{C6H5+} + \ce{C2H2} reaction over a photon energy range of 11.0--15.0 eV, corresponding to maximum internal energies below 3 eV. Panel a displays the \ce{C6H5+} parent ion signal from nitrobenzene. The axis on the top of panel b shows the maximum internal energy deposited in the system, \textit{i.e.}, the difference between the photon energy and the appearance energy of the phenylium ion from ATcT at 300 K. Two primary reaction channels were investigated:
\begin{align}
    \ce{C6H5+} + \ce{C2H2} & \ce{->} & \ce{C8H7+} (m/z\;103)  + h\nu\label{r3}\\
    & \ce{->} & \ce{C8H6+} (m/z\;102) + \ce{H}\label{r4}
\end{align}
Adduct formation (Eq.\ref{r3}) dominates at low internal energy and decreases sharply within 1.5 eV, falling to 5\% of its original reaction cross-section. It remains the main pathway up to 13 eV, above which the loss of H (Eq.\ref{r4}) becomes significant, forming \ce{C8H6+}.

\begin{figure}[!b]
	\includegraphics[width=1\linewidth]{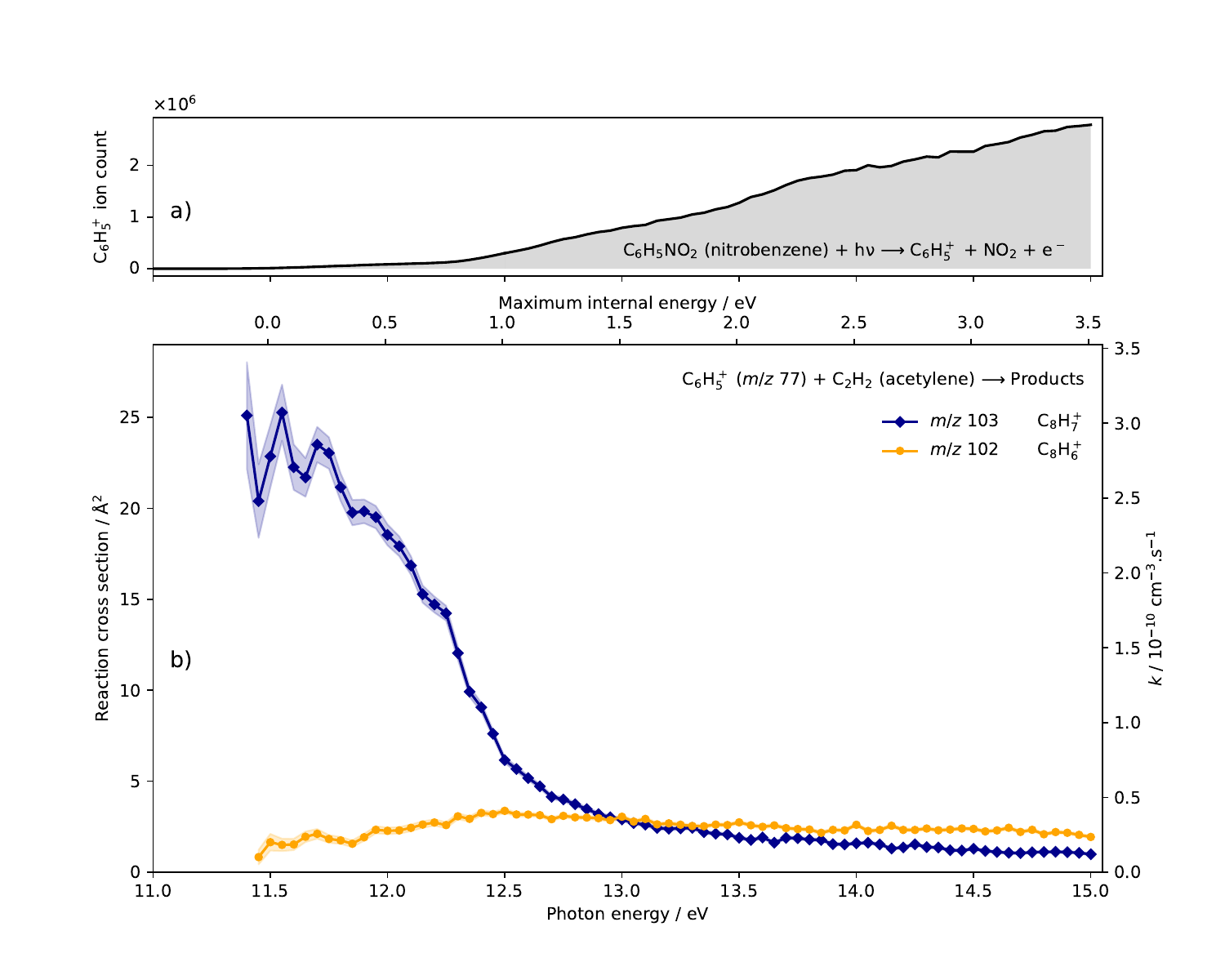}
	\caption{Panel a displays the parent ion signal for the \ce{C6H5+} ion obtained from dissociation ionization of nitrobenzene. Panel b shows the reaction cross sections (left axis) and rate constants (right axis) with \ce{C2H2} as a function of photon energy.} 
	\label{FigC2H2}
\end{figure}

Our experimental findings consistently show that adduct formation is favored, aligning with previous studies \cite{knight1987,anicich2006} where this pathway was often dominant or exclusive. The claim from Kocheril et al. \cite{kocheril2025} that phenylium is unreactive is therefore unexpected and contradicts prior reports of its reactivity with \ce{C2H2} \cite{knight1987,ausloos1989,anicich2006} and \ce{H2}/\ce{D2}\cite{speranza1977,scott1997,ausloos1989,snow1998,ascenzi2003}. Our calculations reaffirm that the entrance channel is barrierless--a hallmark of reactive systems. Pressure may influence the branching ratio between adduct formation and H--loss. In our setup (8.5$\times$ 10$^{-5}$ mbar), pressure is three orders of magnitude higher than in \cite{kocheril2025}, which could enhance stabilization of the adduct (\textit{m/z} 103) over H--loss products (\textit{m/z} 102). However, such pressure differences should affect product distribution rather than eliminate reactivity altogether.  

The most likely explanation for the apparent unreactivity of \ce{C6H5+} reported by Kocheril et al.\cite{kocheril2025} is that, under their low-collision-rate conditions, the species formed in the addition of \ce{C2H2} with protonated diacetylene (\ce{C4H3+}) is not phenylium. The identity of the \ce{C6H5+} isomers present in studies of \ce{C6H5+} + \ce{C2H2} reactions  has never been unambiguously established, for example by spectroscopic means. Instead, the structure of the observed ions is always inferred from the pathway by which \ce{C6H5+} is generated. Two main production routes exist: (i) sequential growth from \ce{C2H2} \cite{brill1981,eyler1984,knight1987,scott1997,kocheril2025}, and (ii) dissociative ionization of benzene and its derivatives \cite{speranza1977,eyler1984,ausloos1989,knight1987,scott1997} and our study. In all reported experiments, ions generated from benzene or its derivatives appear to be more reactive, particularly with \ce{C2H2} and \ce{H2}. Although the experimental observation was always consistent between the different studies, the assignment of the reactive species to phenylium after it was first established in 1977 by Speranza and coworker\cite{speranza1977} was disputed by Eyler and a coworker\cite{eyler1984} who reversed this assignment to phenylium being the unreactive species, an interpretation that was also adopted by Knight et al.\cite{knight1987}. This early debate regarding the reactivity of phenylium was thoroughly documented by Ausloos et al.\cite{ausloos1989}, who presented new evidence highlighting previous misinterpretation. Subsequently, the identification of phenylium as reactive species was further confirmed by additional experimental studies, notably in reactions with \ce{H2}/\ce{D2}\cite{scott1997,ascenzi2003}.

In all studies before ours, \ce{C6H5+} was generated exclusively by energetic ionization methods. Consequently, even when benzene derivatives were used as the precursor, it could not be ensured that phenylium was produced rather than a different isomer. In contrast, our near-threshold dissociative photoionization of nitrobenzene cleanly yields phenylium, since the alternative \ce{C6H5+} isomers are at higher energy. Although indirect, this provides strong evidence that phenylium is indeed reactive toward both \ce{C2H2} and \ce{H2}. In particular, Knight et al. \cite{knight1987} reported a rate of $4 \times 10^{-10}$ cm$^3$·s$^{-1}$ for the reaction of their reactive species with \ce{C2H2}, consistent with our measured value of $3 \times 10^{-10}$ cm$^3$·s$^{-1}$. In contrast, ions formed by sequential addition of \ce{C2H2}, most notably \ce{C4H3+}  + \ce{C2H2}, as in Kocheril et al.\cite{kocheril2025} show no reactivity with \ce{C2H2} or \ce{H2}, implying that phenylium is not produced by this pathway. This conclusion is further supported by theoretical work. Peverati et al. \cite{peverati_2016_insights}, using \textit{ab initio} molecular dynamics (AIMD) simulations, found only limited phenylium formation from the \ce{C4H3+ + C2H2} reaction. Instead, the dominant products were \ce{c-C4H4CCH+} or \ce{C6H4+} + H at higher energies.


The implications of the results of Kocheril \textit{et al.} and our interpretation for astrochemical models are two-fold: i) the formation pathway to phenylium through the reaction \ce{C4H3+} + \ce{C2H2}, should not be considered in models. ii) While the role of phenylium in aromatic formation is not due to a lack of reactivity, its significance might be overestimated. Other routes leading to \ce{C6H7+} (see the results of our astrochemical modelling described in the supplementary material), the primary precursor to benzene, are more efficient. These insights should motivate further theoretical and experimental work to better understand the formation of the first aromatic ring and the crucial impact of isomer-specific reactivity, which remains a major challenge in astrochemical modeling. Addressing this issue is key to resolving the discrepancies between observed abundances of aromatics and predictions from astrochemical models.\cite{byrne2024}

\section*{Methods}

\rev{Reaction pathways were investigated using DFT calculations at the M06-2X/AVTZ level to characterize intermediates, transition states, and long-range interaction potentials. Ion–molecule reactions were studied using the CERISES tandem mass spectrometer coupled to the DESIRS beamline at the SOLEIL synchrotron. \ce{C6H5+} ions were produced by VUV photoionization of nitrobenzene and mass-selected using a quadrupole before entering a reaction cell containing acetylene at 8.5 $\times$ 10$^{-5}$ mbar under single-collision conditions. Ionic products were analyzed with a second quadrupole. Ion internal energy was tuned via photon energy, while collision energy was fixed at 0.15 $\pm$ 0.06 eV. Reaction rate constants were derived from absolute cross sections. }

\section*{Competing interests} 
The authors declare no competing interests.

\section*{Data Availability}
Relevant data supporting the key findings of this study are available within the article. All raw data generated during the current study and scripts used for data reduction are openly available on Zenodo \cite{loison2026data}.

\section*{Author contributions}
 U.J. conceptualized the projects. Co.R., N.S., R.T., C.R., C.A., and U.J. recorded the experimental data. Co.R. performed the data treatment. J-C.L. performed the quantum chemical calculation and astrochemical modelling. U.J. and J-C.L. wrote the manuscript with feedback from all other authors.



\newpage

\begin{center}
{\LARGE Supporting Information: Evidence for Phenylium Reactivity under Interstellar Relevant Conditions
}\\ 
\end{center}
Jean-Christophe Loison,
Corentin Rossi,
Nicolas Solem
Roland Thissen,
Claire Romanzin,
Christian Alcaraz,
Ugo	Jacovella

\bigskip

\begin{figure}[!b]
\begin{center}
	\includegraphics[width=\linewidth]{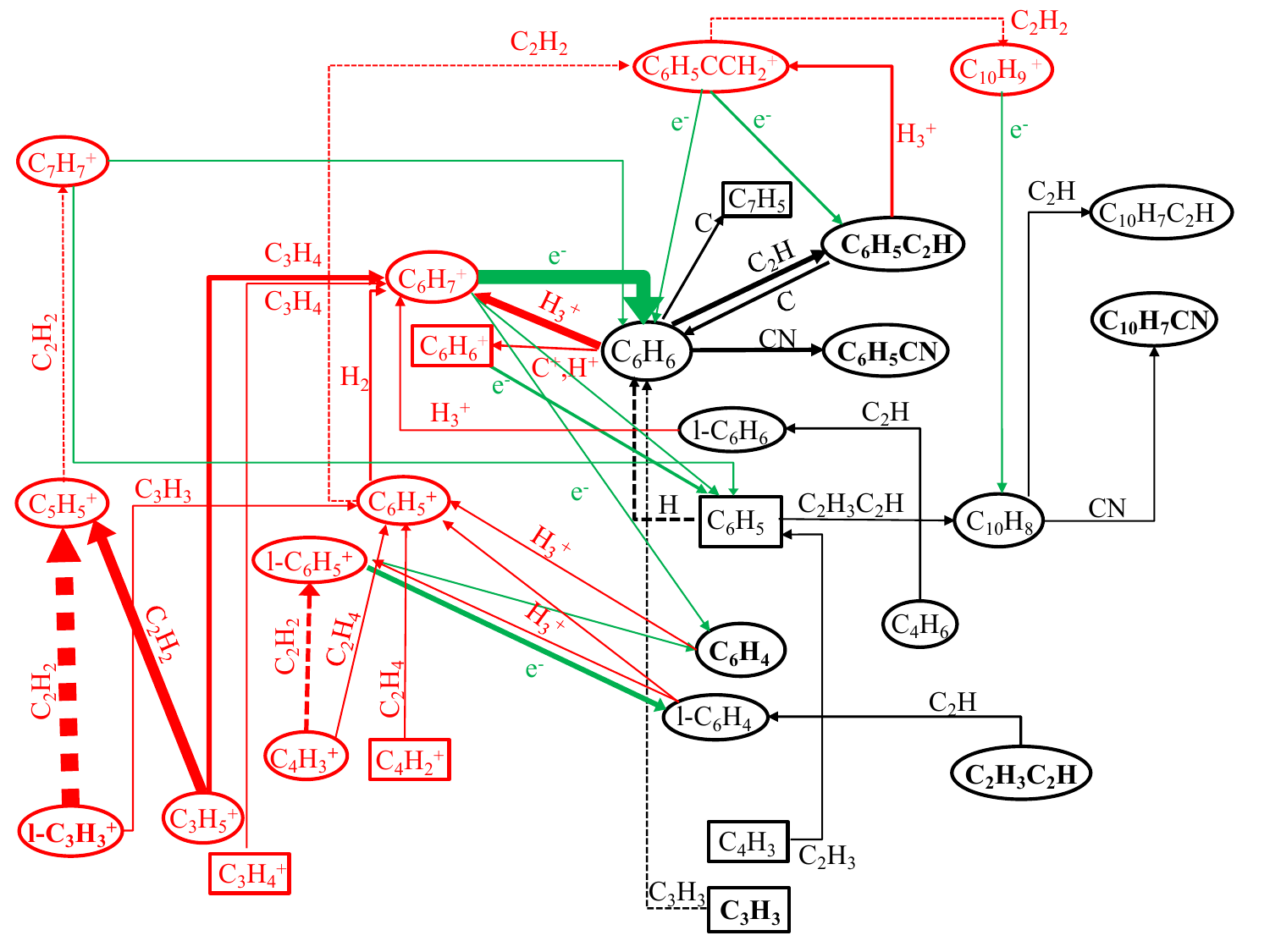}
	\caption{Schematic diagram illustrating key pathways for the formation of aromatic species, with neutral reactions shown in black and ionic reactions in red or green (for Dissociative Re-combination, DR). The thickness of each arrow is proportional to the integrated total production rate, while radiative association reactions are depicted with dashed lines. Radicals are shown within square boxes and close-shell species in circles. Species in bold are those detected in TMC1. The DRs (which are by far the largest fluxes for the reaction of ions that do not react with H$_2$) that do not lead to aromatics are not shown to simplify the figure.}
	\label{flux}
\end{center}
\end{figure}

The main pathways leading to the formation of the first aromatic compounds and subsequent aromatic chemistry are depicted in Supplementary Figure \ref{flux}. A key step in this network is the formation of benzene via the dissociative recombination (DR) of \ce{C6H7+} with an electron. We aimed to assess the significance of our interpretation of the recent findings by Kocheril \textit{et al.} \cite{kocheril2025}, namely that the reaction \ce{C4H3+} + \ce{C2H2} predominantly yields an open-cycle structure of \ce{C6H5+}, denoted as l-\ce{C6H5+}, which does not react with \ce{H2} and therefore does not contribute to the formation of aromatic compounds. To investigate this, we employed the Nautilus code \cite{ruaud_gas_2016}, a three-phase, time-dependent chemical model that includes gas-phase reactions, surface processes on dust grain ices, and bulk chemistry within ice mantles. Our chemical network is based on the kida.uva.2024 database \cite{wakelam_2024}. The network includes 800 species involved in approximately 9000 reactions. Initial elemental abundances assume atomic or ionic forms, with elements ionized if their ionization potential is below 13.6 eV. Simulation parameters are close to those in Table 4 of Hickson \textit{et al.} (2024) \cite{Hickson2024}. Both the grain surface and mantle are chemically active, while accretion and desorption occur only between the surface and gas phase. The dust-to-gas mass ratio is 0.01, and neutral species have a sticking probability of 1. 
Desorption mechanisms include thermal and non-thermal processes (cosmic rays, chemical desorption), with cosmic-ray sputtering \cite{Wakelam2021}. A full description of the surface reaction formalism and simulations is available in Ruaud et al. \cite{ruaud_gas_2016}. The chemical network has been updated to include enhanced aromatic chemistry from previous studies \cite{loison2019,Vuitton2019,Jones2011,Lee2019} as well as new theoretical results from the present study (see Supplementary Table \ref{table_long}). In the updated network, aromatic formation pathways are more diverse, with benzene primarily forming through ionic reactions. Supplementary Figure \ref{flux} summarizes the dominant reactions, where arrow widths indicate the relative reaction fluxes.

\begin{figure}[!t]
\begin{center}
	\includegraphics[width=\linewidth]{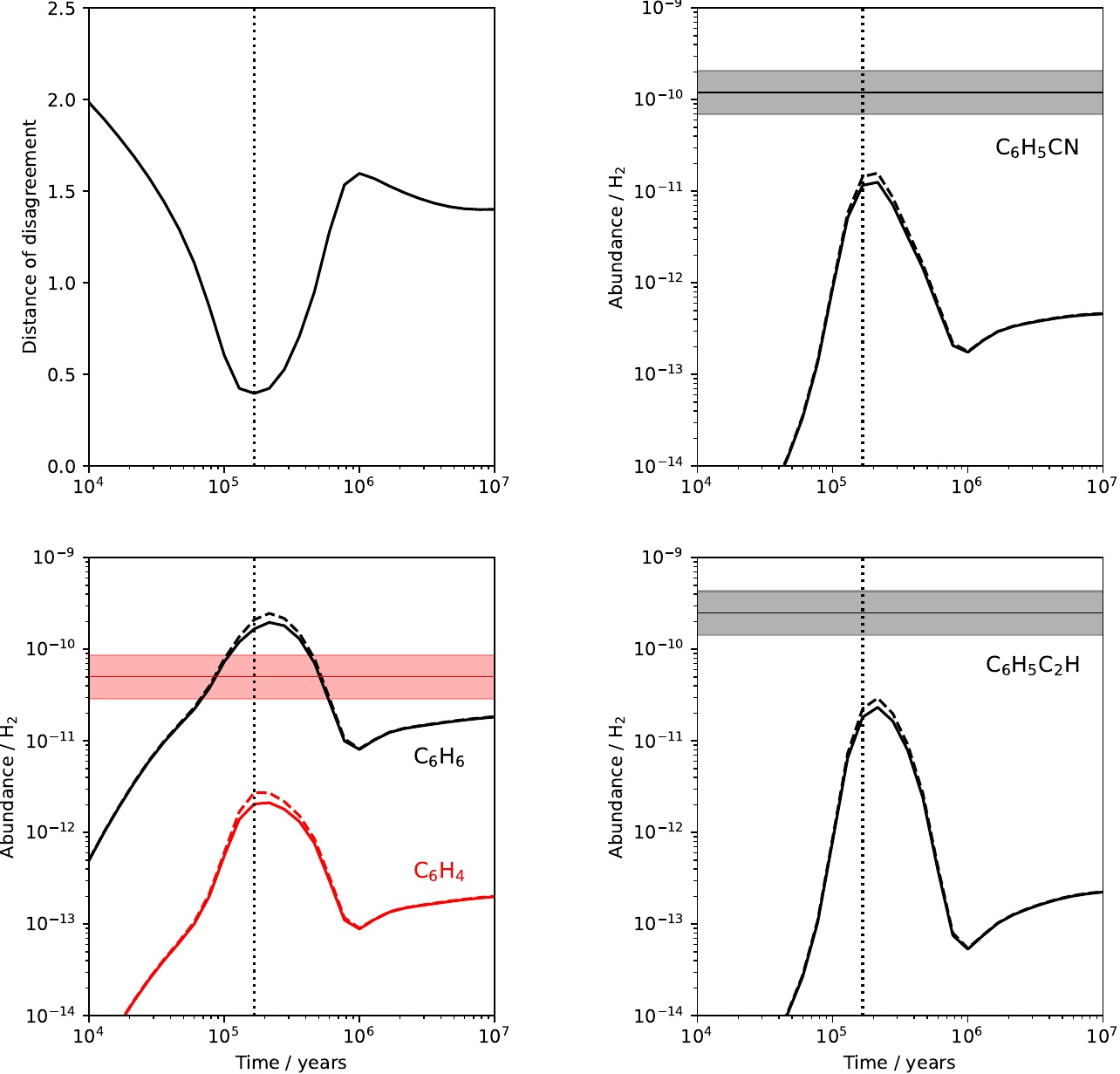}
	\caption{Gas-grain astrochemical model results for the gas-phase formation of \ce{C6H6}, \ce{C6H4}, \ce{C6H5CN}, and \ce{C6H5C2H}. Solid lines represent results from the standard network excluding the gas-phase \ce{C4H3+ + C2H2} reaction leading to phenylium, while dashed lines correspond to the same network with this reaction included. The horizontal rectangles indicate observed abundances of \ce{C6H4}, \ce{C6H5CN}, and \ce{C6H5C2H} in TMC-1 (\cite{Cernicharo2021b}, \cite{mcguire2018a}, \cite{loru2023}), with an estimated uncertainty of $\protect\sqrt{3}$}.
	\label{ab}
\end{center}
\end{figure}

Even though the \ce{C4H3+} + \ce{C2H2} reaction does not produce phenylium, it appears to have only a minor impact on the overall benzene abundance and, by extension, on the formation of other aromatic species. Supplementary Figure \ref{ab} shows the calculated abundances relative to \ce{H2} for \ce{C6H6}, \ce{C6H4}, \ce{C6H5CN}, and \ce{C6H5C2H}, alongside observational data from the TMC-1 molecular cloud \cite{mcguire2018a, Cernicharo2021b, loru2023}. The estimated age of the cloud, marked by a vertical dashed line, corresponds to the time at which our model predictions best match the observed abundances of 67 key species. For comparison, the dotted lines in Supplementary Figure \ref{ab} represent predicted abundances assuming that the \ce{C4H3+} + \ce{C2H2} reaction produces phenylium instead of the non-reactive l-\ce{C6H5+}.

Even under this alternative assumption, the calculated abundances of \ce{C6H4}, \ce{C6H5CN}, and \ce{C6H5C2H} remain one order of magnitude lower than observed values. Comparisons with other models are limited; for example, Byrne et al. \cite{byrne2024} do not report calculated abundances for these species, focusing instead on \ce{C10H7CN}, which their model substantially underestimates.

 For the association reaction \ce{C6H5+} (the reactive isomer) with \ce{H2}, the experimental rate remains nearly constant between 10$^{-6}$ torr \cite{ausloos1989,keheyan2001} and 0.5 torr \cite{snow1998,scott1997,petrie1992}, indicating that the reaction is predominantly governed by the radiative association mechanism. Consequently, the rate under the conditions of the interstellar medium can be estimated \cite{mcewan_new_1999}. Moreover, the reactions of phenylium with \ce{D2} and \ce{H2} have been calculated to proceed without a barrier \cite{ascenzi2003,garcia2009}. Therefore, Phenylium may still play an important role in the broader context of aromatic chemistry, as the reaction with \ce{H2} proceeds at a relatively fast rate, and given the high abundance of \ce{H2} in dense molecular clouds, this pathway is the main bimolecular destruction channel for phenylium and should significantly contribute to the formation of \ce{C6H7+}—a key precursor to benzene. Surprisingly, removing this reaction from the network decreases the predicted benzene abundance by only one order of magnitude. The key reactions leading to the formation of both \ce{C6H7+} and \ce{C6H6}, along with the corresponding reaction rates used in our model, are listed at the end of this document.

This result shows that the formation of benzene is much more varied than previously thought. Moreover, the reactions involved in this formation are not well known. Even the central reaction in our established chemical pathway for benzene formation—the DR of \ce{C6H7+}—remains poorly characterized. Although its rate coefficient is well established \cite{McLain2004, Hamberg2011}, the nature of the reaction products is still uncertain. The only available experimental study \cite{Hamberg2011} suggests that the aromatic ring is preserved, but does not identify the specific products. In our model, we assume that the DR of \ce{C6H7+} predominantly yields benzene, with a minor (5\%) pathway producing \ce{C6H4} + H + \ce{H2}, thereby establishing a potential chemical link between benzene and benzyne in dense molecular clouds.

Interestingly, the KIDA database \cite{wakelam_2014_2015} includes a substantial branching ratio (50\%) for the DR of \ce{C6H7+} yielding \ce{C6H2} + 2\ce{H2} + H, which would result in a net loss of aromaticity. However, our quantum chemical calculations at the M06-2X/AVTZ level indicate that this channel is slightly endothermic (+13 kJ/mol), implying that it is likely inefficient or even non-viable—contrary to the conclusions drawn by Byrne et al. \cite{byrne2024}.

In our model, the dominant pathway for benzene formation is the \ce{C3H5+} + \ce{C3H4} reaction \cite{Anicich2003_Titan}. However, several other reactions also contribute, including \ce{C3H4+} + \ce{C3H4} \cite{Anicich2003_Titan}, l-\ce{C3H3+} + \ce{C3H3}, \ce{C4H2+} + \ce{C2H4} \cite{anicich2006}, and the neutral-neutral reaction \ce{C3H3} + \ce{C3H3}. For the latter, we assume that radiative association efficiently forms benzene—rather than the linear isomers observed at medium and high pressures \cite{Zhao2021, Savee2022,hrodmarsson2024}—and that the complex does not dissociate into \ce{C6H5} + H \cite{Richter2000,Miller2003}.

In contrast to Byrne et al. \cite{byrne2024}, our model indicates that the \ce{C5H2+} + \ce{CH4} reaction plays a negligible role. In the absence of experimental data or even rough estimates, we performed theoretical calculations (assuming the most stable \ce{HC5H+} isomer) and found a barrier of 9 kJ/mol at the M06-2X/AVTZ level for adduct formation, which likely renders this reaction inefficient under low-temperature conditions for producing either \ce{C6H5+} or \ce{C6H4+}.

A major source of uncertainty lies in the chemistry of aromatic compounds, particularly in reaction rates and, even more critically, in branching ratios. The precise isomeric structures formed and their reactivity play a crucial role, especially given the large number of possible isomers for these systems. The most stable isomers are not necessarily the ones preferentially formed, as seen in the case of \ce{C6H5+} isomers produced in the \ce{C4H3+} + \ce{C2H2} reaction. While our results suggest that variations in the \ce{C6H5+} isomer formed via \ce{C4H3+} + \ce{C2H2} do not drastically affect the predicted abundances of aromatic species within our network— which differs significantly from conventional database models.


\begin{table}[h]
 \centering
  \begin{tabular}{cc}
  \hline
  \textbf{Structure} & \textbf{Energy} /kJ.mol$^{-1}$\\
  \hline
  \ce{C6H5+ + C2H2} & 0 \\
  \textbf{(1)} & -275.7 \\
  \textit{TS$_{\text{1-2}}$} & -179.3 \\
  \textit{TS$_{\text{1-3}}$} & -218.8 \\
  \textbf{(2)} & -346.2 \\
  \textbf{(3)} & -263.3 \\
  \textit{TS$_{\text{2-6}}$} & -119.8 \\
  \textit{TS$_{\text{3-5}}$} & -58.8 \\
  \textbf{(4)} &  -149.3 \\ 
  \textit{TS$_{\text{4-5}}$} & -59.2 \\
  \textbf{(i)} & 6.8 \\
  \textbf{(5)} & -197.1 \\
  \textit{TS$_{\text{5-6}}$} & -39.9 \\
  \textbf{(ii)} & 15.1 \\
  \textbf{(6)} & -309.6 \\
  \textbf{(iii)} & -24.1 \\
  \hline
  \end{tabular}
  \caption{Calculated energies for the minima and transition states of the potential energy surface for the \ce{C6H5+} + \ce{C2H2} reaction. Energies are given relatively to the energy of the two reactants. All calculations were performed using M06-2X/AVTZ level.}
  \label{table_small}
\end{table}

{\small
\begin{longtable}{rlcccl}
\caption{Chemical network used for the model. Rate coefficients $\alpha$($T/300)^{\beta}$exp(-$\gamma$/$T$) are in units of $\mathrm{cm^3\,s^{-1}}$. }\\
\hline \hline
\# & Reaction & $\alpha$ & $\beta$ & $\gamma$ & ref \\
\hline
\endfirsthead
\caption{continued.}\\
\hline \hline
\# & Reaction & $\alpha$ & $\beta$ & $\gamma$ & ref \\
\hline
\endhead
\hline
\endfoot
\rule{0pt}{3ex} 
1 & \ce{H3+ + C6H4} $\rightarrow$  \ce{C6H5+ + H2} & 3.9E-09 & -0.5 & 0 &/ \ce{H3+ + C6H6}\\
\\
2 & \ce{H3+} + l-\ce{C6H4} $\rightarrow$  \ce{C6H5+ + H2} & 3.9E-09 & 0 & 0 &/ \ce{H3+ + C6H6}\\
\\
3 & \ce{H3+ + C6H6} $\rightarrow$  \ce{C6H7+ + H2} & 3.9E-09 & 0 & 0 &\cite{milligan2002}\\
\\
4 & \ce{H3+} + l-\ce{C6H6}$\rightarrow$  l-\ce{C3H3+ + CH3CCH + H2} & 3.9E-09 & -0.5 & 0 &/ \ce{H3+ + C6H6}\\
\\
5 & \ce{H3+ + C7H5} $\rightarrow$  \ce{C6H6+ + CH2} & 3.9E-09 & 0 & 0 &/ \ce{H3+ + C6H6}\\
\\
6 & l-\ce{C3H3+} + l-\ce{C3H} $\rightarrow$  \ce{C6H2+ + H2} & 1.0E-09 & -0.5 & 0 &
\multirow{2}{4.4cm}{/ l-\ce{C3H3+ + C2H2}, l-\ce{C3H3+ + C2H4} \cite{Anicich2003}}\\
& \hspace*{3.0cm} \ce{C6H3+ + H} & 1.0E-09 & -0.5 & 0 &\\
\\
7 & \ce{C3H4+ + CH3CCH}  $\rightarrow$ \ce{C6H7+ + H} & 7.5E-10 & -0.5 & 0 &
\multirow{1}{4cm}{\cite{Anicich2003, anicich2003b, mcEwan2007}.}\\
 & \hspace*{3.2cm} \ce{C3H5+ + C3H3} & 3.5E-10 & -0.5 & 0 &\\
\\
8 & \ce{C3H4+ + CH2CCH2}  $\rightarrow$ \ce{C6H7+ + H} & 8.0E-10 & 0 & 0 &
\multirow{1}{4cm}{\cite{anicich_evaluated_1993, anicich2003b}.}\\
 & \hspace*{3.2cm} \ce{C6H6+ + H} & 3.0E-10 & 0 & 0 &\\
\\
9 & \ce{C3H5+ + CH3CCH}  $\rightarrow$ \ce{C6H7+ + H2} & 7.0E-10 & -0.5 & 0 & \cite{anicich2006}\\ 
\\
10 & \ce{C3H5+ + CH2CCH2}  $\rightarrow$ \ce{C6H7+ + H2} & 7.0E-10 & 0 & 0 &/ \ce{C3H5+ + CH3CCH} \\ 
\\
11 & \ce{C4H2+ + C2H4}  $\rightarrow$ \ce{C6H5+ + H} & 8.0E-10 & 0 & 0 &\cite{anicich2006}.\\
 & \hspace*{2.6cm} \ce{C6H4+ + H2} & 2.0E-10 & 0 & 0 &\\
\\
12 & \ce{C4H3+ + C2H2}  $\rightarrow$ l-\ce{C6H5+ + h\nu} & 2.2E-10 & 0 & 0 &
\multirow{1}{4cm}{\cite{Anicich2003,anicich2006,knight1987,kocheril2025,peverati2016}}\\
\\
13 & \ce{C4H3+ + C2H4}  $\rightarrow$ \ce{C6H5+ + H2} & 1.2E-10 & 0 & 0 &
\multirow{2}{4cm}{\cite{mcEwan2007} \ce{C6H5+} is likely the phenylium}\\
\\
\\
14 & \ce{C5H5+ + C2H2}  $\rightarrow$ \ce{C6H7+ + h\nu} & 1.6E-11 & -0.5 & 0 & 
\multirow{1}{4cm}{\cite{anicich2006, mcEwan2007, Ozturk1989}}\\
& \hspace*{2.65cm} \ce{C6H5CH2+ + h\nu} & 1.6E-11 & -0.5 & 0 &\\
\\
15 & \ce{C6H5+ + H2}  $\rightarrow$ \ce{C6H7+ + h\nu} & 6.0E-11 & 0 & 0 &
 \multirow{1}{4cm}{\cite{ausloos1989, keheyan2001, McEwan1999, petrie1992, scott1997, snow1998}}\\
 \\
16 & \ce{N2H+ + C6H6}  $\rightarrow$ \ce{C6H7+ + N2} & 1.6E-09 & 0 & 0 & \cite{wakelam_2024}\\
\\
17 & \ce{HCO+ + C6H4}  $\rightarrow$ \ce{C6H5+ + CO} & 1.6E-09 & 0 & 0 &/ \ce{HCO+ + C6H6}\\
\\
18 & \ce{HCO+} + l-\ce{C6H4}  $\rightarrow$ \ce{C6H5+ + CO} & 1.6E-09 & -0.5 & 0 &/ \ce{HCO+ + C6H6}\\
\\
19 & \ce{HCO+ + C6H6}  $\rightarrow$ \ce{C6H7+ + CO} & 1.6E-09 & 0 & 0 &\cite{wakelam_2024}\\
\\
20 & \ce{H + C6H5}  $\rightarrow$ \ce{C6H6 + h\nu} & 1.0E-10 & 0 & 0 &
\multirow{1}{4cm}{\cite{Ackermann1990,Vuitton2012}}\\
\\
21 & \ce{C + C6H6}  $\rightarrow$ \ce{C7H5 + H} & 3.0E-10 & 0 & 0 & \cite{bergeat2001}\\
\\
22& \ce{C + C6H5C2H}  $\rightarrow$  \ce{C6H6 + C3} & 3.0E-10 & 0 & 0 &/ \ce{C + C6H6}.\\
\\
23 & \ce{C2H + C2H3C2H}  $\rightarrow$ l-\ce{C6H4 + H} & 3.0E-10 & 0 & 0 &
\multirow{2}{4cm}{/ \ce{C2H + C4H2} \cite{Landera2008} and \ce{C2H + C4H6} \cite{Jones2011}}\\
\\
\\
24& \ce{C2H + CH2CHCHCH2}  $\rightarrow$  l-\ce{C6H6 +H} & 2.0E-10 & 0 & 0 &\cite{Jones2011}.\\
\\
25& \ce{C2H + C6H6}  $\rightarrow$ \ce{C6H5C2H + H} & 3.28E-10 & -0.18 & 0 &
\multirow{1}{4cm}{\cite{Goulay2006, woon_stability_1996}.}\\
\\
26& \ce{C2H3 + C4H3}  $\rightarrow$ \ce{C6H5 + H} & 4.0E-11 & 0 & 0 &
\multirow{2}{4cm}{\cite{Duran1988} and \cite{Reis2024}, linear l-\ce{C6H5} isomers may also be produced.}\\
\\
\\
\\
27 & \ce{C3H3 + C3H3}  $\rightarrow$ \ce{C6H6 + h\nu} & 6.5E-11 & 0 & 0 &
\multirow{2}{4cm}{\cite{Miller2003,Vuitton2019, hrodmarsson2024}. Other \ce{C6H6} isomers may also be produced and the \ce{C6H5} + H pathway may not be negligible.}\\
\\
\\
\\
\\
\\
28& \ce{CN + C6H6}  $\rightarrow$ \ce{C6H5CN + H} & 4.0E-10 & 0 & 0 &\cite{Trevitt2009}.\\
\\
29& \ce{C6H5+ + e-}  $\rightarrow$ \ce{C6H4 + H}  & 2.0E-07 & -0.3 & 0 &
\multirow{1}{4cm}{rate from \cite{Fournier2013}}\\
& \hspace*{2.2cm} \ce{C6H2 + H + H2} &  1.0E-06 & -0.3 &  &\\
\\
30& l-\ce{C6H5+ + e-}  $\rightarrow$ \ce{C6H4 + H}  & 1.0E-06 & -0.3 & 0 &/ \ce{C6H5+ + e-} \\
& \hspace*{2.4cm} \ce{C6H2 + H + H2}  & 1.0E-06 & -0.3 & 0 &\\
\\
31& \ce{C6H6+ + e-}  $\rightarrow$ \ce{C6H5 + H} & 1.0E-06 & -0.69 & 0 &
\multirow{2}{4cm}{rate constant from \cite{Hamberg2011}, branching ratio guessed by comparison with \ce{C6H6} photodissociation}\\
& \hspace*{2.2cm} \ce{C6H4 + H2}  & 1.0E-07 & -0.69 & 0 &\\
& \hspace*{2.2cm} \ce{C3H3 + C3H3} & 1.0E-07 & -0.69 & 0 &\\
\\
\\
32& \ce{C6H7+ + e-}  $\rightarrow$ \ce{C6H6 + H}  & 2.0E-06 & -0.83 & 0 &
\multirow{1}{4cm}{Global rate from \cite{Hamberg2011}}\\
& \hspace*{2.2cm} \ce{C6H4 + H + H2}  & 1.0E-07 & -0.83 & 0 &\\
& \hspace*{2.2cm} \ce{C6H5 + H + H} & 1.0E-07 & -0.83 & 0 &\\
\\
33& \ce{C6H5CH2+ + e-}  $\rightarrow$ \ce{C6H6 + CH2}  & 3.0E-07 & -0.70 & 0 &
\multirow{2}{4cm}{rate from \cite{Fournier2013}, arbitrary branching ratios}\\
& \hspace*{2.9cm} \ce{C6H6 + CH}  & 3.0E-07 & -0.70 & 0 &\\
& \hspace*{2.9cm} \ce{CH3CCH + C4H2 + H} & 1.0E-07 & -0.70 & 0 &\\
& \hspace*{2.9cm} \ce{CH2CCH2 + C4H2 + H} & 1.0E-07 & -0.70 & 0 &\\
\\
\\
34& \ce{C7H7++ + e-}  $\rightarrow$ \ce{C6H6 + CH2}  & 3.0E-07 & -0.70 & 0 &
\multirow{2}{4cm}{rate from \cite{Fournier2013}, arbitrary branching ratios}\\
& \hspace*{2.45cm} \ce{C6H6 + CH}  & 3.0E-07 & -0.70 & 0 &\\
& \hspace*{2.45cm} \ce{CH3CCH + C4H2 + H} & 1.0E-07 & -0.70 & 0 &\\
& \hspace*{2.45cm} \ce{CH2CCH2 + C4H2 + H} & 1.0E-07 & -0.70 & 0 &\\
\\
35& \ce{C6H5CCH2+ + e-}  $\rightarrow$ \ce{C6H5C2H + H}  & 1.0E-06 & -0.83 & 0 &
\multirow{2}{4cm}{/ \ce{C6H6+ + e-}, \ce{C6H5CH2+ + e-} and \ce{C7H7+ + e-}. Large uncertainties, \ce{C6H6 + C2H} and \ce{C6H6 + C2H2} may also be produced}\\
\\
\\
\\
\\
\\
\\
36& \ce{C6H5CNH+ + e-}  $\rightarrow$ \ce{C6H5CN + H}  & 1.5E-06 & -0.8 & 0 &
\multirow{2}{4cm}{/ \ce{C6H6+ + e-}, \ce{C6H5CH2+ + e-}, \ce{C7H7+ + e-} and \ce{HC3NH+ + e-}}\\
& \hspace*{2.9cm} \ce{C6H5 + HCN} & 2.5E-07 & -0.8 & 0 &\\
& \hspace*{2.9cm} \ce{C6H5 + HNC}  & 2.5E-07 & -0.8 & 0 &\\

\label{table_long}
\end{longtable}
}

\end{document}